\documentclass[nohyper,12pt,letterpaper]{JHEP3}
\usepackage{epsfig}
\usepackage[latin1]{inputenc}
\usepackage{bbm,amsfonts}
\usepackage{graphicx}
\usepackage{amssymb,amsmath}

\author{Marco S. Bianchi$^\ast$ and
  Matias Leoni$^{\dag}$\\\\
$^\ast$Departamento de F\'isica, Universidad de Oviedo, 33007, Oviedo, Spain
  \\\\
  $^{\dag}$Physics Department, FCEyN-UBA \& IFIBA-CONICET\
  Ciudad Universitaria, Pabell\'on I, 1428, Buenos Aires, Argentina
  \qquad\\\\
  E-mail: \email{bianchimarco@uniovi.es, leoni@df.uba.ar
  }
}

\abstract{We compute the four--point amplitude in $3d$ ${\cal N}=8$ SYM at two loops, by solving the three dimensional scalar doublebox in dimensional regularization. We compare it to the same result in the ABJM theory, to which maximal SYM should flow in the infrared at strong coupling. After proper rescalings, we find that the two amplitudes coincide in the Regge limit.}

\preprint{October 2012 \\ FPAUO-12/14}

\title{$\mathcal{N}=8$ SYM vs. $\mathcal{N}=6$ Chern-Simons: Four--point amplitudes at two--loops}


\csname @addtoreset\endcsname{equation}{section}

\def\bseq{\begin{subequation}}  % = 1a 1b
\def\eseq{\end{subequation}}
\def\bsea{\begin{subeqnarray}}  % = 1.1a 1.1b
\def\esea{\end{subeqnarray}}
\newcommand{\beq}{\begin{equation}}
\newcommand{\bea}{\begin{eqnarray}}
\newcommand{\eea}{\end{eqnarray}}
\newcommand{\eeq}{\end{equation}}

\newcommand {\non}{\nonumber}

\newcommand{\g}{\gamma}

\newcommand{\e}{\epsilon}

\renewcommand{\l}{\lambda}

\newcommand{\p}{\pi}
\renewcommand{\P}{\Pi}

\def\Mb{\kern 2pt\mathchoice
        {%displaystyle
         \vbox{\hrule width10pt height 0.4pt depth 0pt
         \kern 1.2pt\hbox{\kern -2pt$\displaystyle M$}}}
        {%textstyle
         \vbox{\hrule width10pt height 0.4pt depth 0pt
         \kern 1.2pt\hbox{\kern -2pt$\textstyle M$}}}
        {%scriptstyle \kern 0.5pt
\vbox{\hrule width6pt height 0.4pt depth 0pt
         \kern 1.0pt\hbox{\kern -2pt$\scriptstyle M$}}}
        {%scriptscriptstyle \kern 0.5pt
         \vbox{\hrule width5pt height 0.4pt depth 0pt
         \kern 0.8pt\hbox{\kern -2pt$\scriptscriptstyle M$}}}}

\def\Sb{\kern 2pt\mathchoice
        {%displaystyle
         \vbox{\hrule width6pt height 0.4pt depth 0pt
         \kern 1.2pt\hbox{\kern -2pt$\displaystyle S$}}}
        {%textstyle
         \vbox{\hrule width6pt height 0.4pt depth 0pt
         \kern 1.2pt\hbox{\kern -2pt$\textstyle S$}}}
        {%scriptstyle
         \vbox{\hrule width3.5pt height 0.4pt depth 0pt
         \kern 1.0pt\hbox{\kern -2pt$\scriptstyle S$}}}
        {%scriptscriptstyle
         \vbox{\hrule width3pt height 0.4pt depth 0pt
         \kern 0.8pt\hbox{\kern -2pt$\scriptscriptstyle S$}}}}

\def\Rb{\kern 2pt\mathchoice
        {%displaystyle
         \vbox{\hrule width5.5pt height 0.4pt depth 0pt
         \kern 1.2pt\hbox{\kern -2.5pt$\displaystyle R$}}}
        {%textstyle
         \vbox{\hrule width5.5pt height 0.4pt depth 0pt
         \kern 1.2pt\hbox{\kern -2.5pt$\textstyle R$}}}
        {%scriptstyle
         \vbox{\hrule width3.5pt height 0.4pt depth 0pt
         \kern 1.0pt\hbox{\kern -2.2pt$\scriptstyle R$}}}
        {%scriptscriptstyle
         \vbox{\hrule width3pt height 0.4pt depth 0pt
         \kern 0.8pt\hbox{\kern -2.2pt$\scriptscriptstyle R$}}}}

  \def\pp{{\mathchoice
      {%displaystyle
          \kern 1pt%
          \raise 1pt
          \vbox{\hrule width5pt height0.4pt depth0pt
            \kern -2pt
            \hbox{\kern 2.3pt
              \vrule width0.4pt height6pt depth0pt
              }
            \kern -2pt
            \hrule width5pt height0.4pt depth0pt}%
            \kern 1pt
       }
        {%textstyle
          \kern 1pt%
          \raise 1pt
          \vbox{\hrule width4.3pt height0.4pt depth0pt
            \kern -1.8pt
            \hbox{\kern 1.95pt
              \vrule width0.4pt height5.4pt depth0pt
              }
            \kern -1.8pt
            \hrule width4.3pt height0.4pt depth0pt}%
            \kern 1pt
        }
        {%scriptstyle
          \kern 0.5pt%
          \raise 1pt
          \vbox{\hrule width4.0pt height0.3pt depth0pt
            \kern -1.9pt  %[e]=0.15pt
            \hbox{\kern 1.85pt
              \vrule width0.3pt height5.7pt depth0pt
              }
            \kern -1.9pt
            \hrule width4.0pt height0.3pt depth0pt}%
            \kern 0.5pt
        }
        {%scriptscriptstyle
          \kern 0.5pt%
          \raise 1pt
          \vbox{\hrule width3.6pt height0.3pt depth0pt
            \kern -1.5pt
            \hbox{\kern 1.65pt
              \vrule width0.3pt height4.5pt depth0pt
              }
            \kern -1.5pt
            \hrule width3.6pt height0.3pt depth0pt}%
            \kern 0.5pt%}
        }
    }}

  \def\mm{{\mathchoice
               {%displaystyle
                 \kern 1pt
           \raise 1pt    \vbox{\hrule width5pt height0.4pt depth0pt
                  \kern 2pt
                  \hrule width5pt height0.4pt depth0pt}
                 \kern 1pt}
               {%textstyle
                \kern 1pt
           \raise 1pt \vbox{\hrule width4.3pt height0.4pt depth0pt
                  \kern 1.8pt
                  \hrule width4.3pt height0.4pt depth0pt}
                 \kern 1pt}
               {%scriptstyle
                \kern 0.5pt
           \raise 1pt
                \vbox{\hrule width4.0pt height0.3pt depth0pt
                  \kern 1.9pt
                  \hrule width4.0pt height0.3pt depth0pt}
                \kern 1pt}
               {%scriptscriptstyle
               \kern 0.5pt
         \raise 1pt  \vbox{\hrule width3.6pt height0.3pt depth0pt
                  \kern 1.5pt
                  \hrule width3.6pt height0.3pt depth0pt}
               \kern 0.5pt}
               }}

\def\pd{{\kern0.5pt
           + \kern-5.05pt \raise5.8pt\hbox{$\textstyle.$}\kern
0.5pt}}

\def\pmd{{\kern0.5pt
          \pm \kern-5.05pt
\raise6.3pt\hbox{$\textstyle.$}\kern1.5pt}}

\def\md{{\mathchoice
   {%displaystyle
      {{\kern 1pt - \kern-6.2pt \raise5pt\hbox{$\textstyle.$}\kern
1pt}}}
    {%textstyle
      {{\kern 1pt - \kern-6.2pt \raise5pt\hbox{$\textstyle.$}\kern
1pt}}}
    {%scriptstyle
      {\kern0.5pt - \kern-5.05pt
\raise3.4pt\hbox{$\textstyle.$}\kern0.5pt}}
    {%scriptscriptstyle
      {\kern0.5pt - \kern-5.05pt
\raise3.4pt\hbox{$\textstyle.$}\kern0.5pt}}}}

\def\beq{\begin{equation}}
\def\eeq{\end{equation}}
\def\bea{\begin{eqnarray}}
\def\eea{\end{eqnarray}}

\def\g{\gamma}

\def\e{\epsilon}

\def\l{\lambda}

\begin{document}

\section{Introduction.}

Amazing progress in the computation of scattering amplitudes and in a deeper understanding of their properties has been attained in the last years. Most of these advances have been achieved in the realm of ${\cal N}=4$ SYM theory in four dimensions, however, in order to gain a broader insight on the subject, it is worthwhile exploring amplitudes in other contexts. Since the realization of the ABJM model \cite{ABJM,ABJ} in three dimensions, an exciting new environment where one can study amplitudes was opened. Many results are now already available for ${\cal N}=6$ Chern--Simons matter theories: after the first investigation of \cite{ABL}, the structure of tree level amplitudes was analyzed  \cite{BLM, HL}, discovering an underlying dual superconformal symmetry \cite{Drummond:2008vq}, which combined with the original superconformal invariance of the model, gives rise to a Yangian \cite{DHP}. This symmetry has been proven to extend to all tree level amplitudes and loop integrands \cite{Gang:2010gy}, by virtue of a three dimensional version of recursive relations \cite{BCF,BCFW}, and was exploited to propose an orthogonal Grassmannian integral formula \cite{Lee}, in close relationship with the ${\cal N}=4$ one \cite{ArkaniHamed:2009dn}.

While this appears to be a symmetry at weak coupling, a neat dual interpretation, like the one of \cite{BM,Beisert:2008iq} in four dimensions, is still lacking \cite{ADO}--\cite{Colgain:2012ca}.

At higher order, amplitudes have been computed at one loop through unitarity \cite{Bargheer:2012cp}--\cite{Brandhuber:2012wy} and directly through Feynman supergraphs \cite{Bianchi:2012cq}. Such amplitudes are IR finite and show a three dimensional version of the holomorphic anomaly \cite{Cachazo:2004by}.

At two loops the four--point amplitude was computed \cite{CH}--\cite{BLMPS2}. It exhibits dual conformal invariance \cite{Drummond:2006rz}, duality with the bosonic four--cusped Wilson loop \cite{Drummond:2007cf} computed in \cite{HPW} and there are hints that it may enjoy exponentiation, similarly to the BDS four dimensional formula \cite{BDS}.
The two--loop four--point amplitude in ABJM strikingly resembles its one--loop analogue in four dimensional ${\cal N}=4$ SYM, and a relation connecting the two, to all orders in the dimensional regularization parameter expansion was determined in \cite{BLP}, which might potentially extend to all loops, if the amplitudes do really exponentiate.

Furthermore the computation of the six--point amplitude at two--loops has been recently carried out \cite{yutin}.\\

Apart from Chern--Simons theory, amplitudes have been studied in three dimensions for supersymmetric Yang--Mills models.
The Yang--Mills action being superrenormalizable and having a dimensionful coupling, prevents the theory from being superconformal, and therefore from having an AdS/CFT dual. Hence one may expect that the nice properties of ABJM amplitudes may not be shared by SYM ones.
Still, these theories may reveal interesting aspects, since they are expected to flow to Chern--Simons matter theories in the infrared.

In particular maximal supersymmetric Yang--Mills theory with gauge group $U(N)$, which describes the low--energy worldvolume theory of a stack of $N$ D2--branes, is believed to have ABJM $U(N)_{1}\times U(N)_{-1}$ as a conformal fixed point of its RG flow in the deep infrared. In this phase the analysis is hampered by the inherently strongly coupled regime of the two theories, nevertheless evidence of their equivalence has been pointed out in \cite{Kapustin:2010xq}--\cite{Gang:2011xp}, matching their partition functions and superconformal indices.

It is not clear how this may reflect on the structure of scattering amplitudes, however \cite{Agarwal:2011tz} showed how the $SO(8)$ R--symmetry expected for ${\cal N}=8$ enhanced ABJM with level one can be recovered at four points in maximally SYM scattering, and such an analysis was extended in \cite{Agarwal:2012jj}.

By analyzing tree level amplitudes of ${\cal N} = 8$ SYM in on--shell superspace,
and by means of recursion relations, \cite{Lipstein} proved that they enjoy dual
conformal covariance. This also applies to the integrands of loop
amplitudes by means of unitarity. Furthermore, that paper developed a
prescription to go from four dimensional SYM amplitudes to three
dimensional ones by means of dimensional reduction, and used it to show
that ${\cal N} = 8$ SYM amplitudes in three dimensions have helicity structure and that beyond
four--points, the amplitudes should have $SO(7)$ R--symmetry. 

In the same
paper, the authors also argued that in maximal three dimensional SYM, the loop integrands
can be obtained from four dimensional ${\cal N}=4$ SYM and subsequently showed that at one loop,
all MHV amplitudes vanish and all non--MHV amplitudes are finite. This is
at least consistent with what occurs in ABJM. The authors in \cite{Lipstein} also
point out that at two loops the four--point amplitudes of ABJM and ${\cal N}=8$
SYM might show some similarity, motivated by their conjectured relation
and by the fact that they share dual conformal
covariance.

In this short note we explore this possibility by explicitly computing the two--loop ratio of the four--point amplitude to its tree level counterpart for ${\cal N}=8$ SYM in three dimensions.

Following dimensional reduction, this task may be accomplished by computing the three dimensional scalar doublebox.
As expected, this integral suffers from infrared divergences which we regulate through dimensional regularization.
This computation is described in section \ref{sec:integral}, where we also give the final expression for the amplitude.

In section \ref{sec:properties} we comment on the outcome.
Given the dimensionful nature of the coupling constant, we expect the ratio to have mass dimension $-2$ at two--loops, which is indeed the case, due to an overall $\left(s^{-1}+t^{-1}\right)$ factor.
The amplitude presents infrared singularities, in the form of $\e^{-2}$ poles, whereas subleading divergences may be absorbed into a proper rescaling of the mass parameter of dimensional regularization.
The finite part is made of a squared logarithm of the ratio of Mandelstam invariants, a subleading logarithm multiplied by an $s\leftrightarrow t$ antisymmetric combination, and a constant piece.
In contrast to ABJM, this amplitude does not exhibit uniform transcendentality.

Since the theory does not even possess conformal invariance one suspects that a duality with light--like Wilson loops should not occur.
We verify that this suspicion is correct by computing the first order correction to the four cusped Wilson--loop, which is non--vanishing in contrast to the result for the amplitude of \cite{Lipstein}. At two loops we also check that not even the UV divergences of the cusp in the light--like Wilson loop match the IR singularities of the amplitude.
Actually at two loops the Wilson loop only shows $\e^{-1}$ UV poles, as a result of milder short distance singularity, since the theory is superrenormalizable.

In the last section we compare the SYM amplitude with the ABJM one.
Upon redefining the relative coupling constants in such a way that the effective YM parameter is dimensionless, and that the coefficients in front of the infrared divergent piece coincide, the two amplitudes exhibit some partial resemblance, spoiled by the presence of the term proportional to $\log{s/t}$, absent in the ABJM case. Quite interestingly, the maximal transcendentality part of the constant coincides with that of ABJM.
Taking the Regge limit $s/t \rightarrow 0$, the leading logarithm approximation reproduces exactly the ABJM case, in the sense that the $\log^2 s/t$ coefficient is precisely the same. For ABJM, thanks to the duality with the Wilson loop, an anomalous conformal Ward identity \cite{Drummond:2007aua,Drummond:2007au} fixes this coefficient to be one half that multiplying the $-\e^{-2}$ poles.

\section{Computation of the amplitude: the three dimensional scalar doublebox.}\label{sec:integral}

We consider ${\cal N}=8$ SYM theory in three dimensions, with unitary gauge group $U(N)$, whose rank we take large in the planar limit $N\gg 1$, and coupling constant $g_{YM}$, having mass dimension $1/2$.

At weak coupling, and in the planar limit, we expand the color ordered four gluon amplitude in the power series
\begin{equation}
\mathcal{A}_{4} = \mathcal{A}_{4}^{(0)}\, \sum_{L=0}^{\infty}\, \left(\frac{2\, g^2_{YM} N \, e^{-\g_E\e} \mu^{2\e}}{(4\pi)^{3/2-\e}}\right)^L\, {\cal M}_4^{(L)}
\end{equation}
where $\e = \frac{3-d}{2}$ is the dimensional regularization parameter and $\mu$ a mass scale.

We want to compute the two--loop correction to ${\cal M}_4$.
Following the derivation in \cite{Lipstein}, this amounts to borrowing the four dimensional result for the integrand and shifting the dimension of the Feynman integrals to $d=3-2\e$.
This is motivated by the observation that the integrands of three dimensional ${\cal N}=8$ SYM obey the same transformation properties of those of ${\cal N}=4$ SYM in four dimensions, under dual inversion, and are thus the same.

Therefore we compute the three dimensional scalar massless doublebox\footnote{We follow the conventions of \cite{BDS} for the normalization of the integral.} in Fig. \ref{fig:doublebox}.
\begin{equation}\label{eq:intnorma}
\mathcal{I}_{4s}^{(2)} = - e^{2\g_E\e}\, \p^{-3+2\e}\, \int\, \frac{d^{3-2\e} k\, d^{3-2\e} l}{k^2 (k-p_2)^2 (k+p_1)^2 (k+l+p_1+p_4)^2 (l+p_4)^2 (l-p_3)^2 l^2}
\end{equation}
\FIGURE{
  \centering
\includegraphics[width = .65 \textwidth]{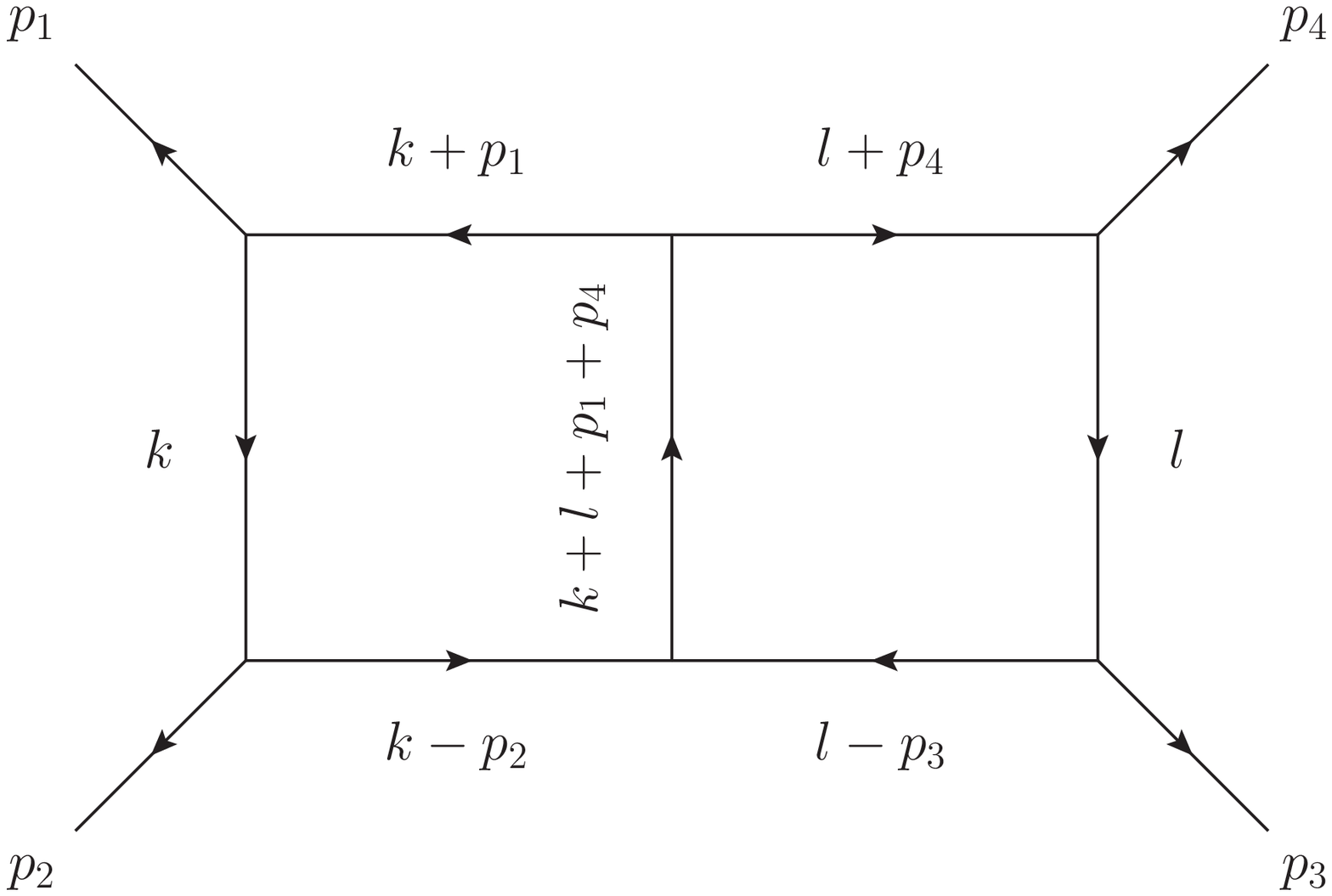}
\caption{The scalar doublebox.}
\label{fig:doublebox}
}
Note that this integral is not dual conformally invariant, a property which is prevented because we have lowered the dimension of the integration measure to three. However it inherits dual conformal covariance from four dimensions.
Calling $s/t \equiv x$, its Mellin--Barnes representation reads
\begin{equation}
\mathcal{I}_{4s}^{(2)} = -\frac{e^{2\g_E \e}}{(-s)^{4+2\e}}\, f\left( x , \e \right)
\end{equation}
where
\begin{align}\label{MBRep}
f\left( x , \e \right) = \, & \frac{1}{\Gamma (-2 \epsilon -1) }\, \int_{-i\infty}^{+i\infty}\, \prod_{i=1}^{4} \frac{dz_i}{2\p i} \, \Gamma \left(-z_1\right) \Gamma \left(z_1+1\right) \Gamma \left(z_1-z_4+1\right) \times \non\\&  \Gamma \left(-z_2-z_3-z_4\right)  \Gamma \left(z_2+z_4\right) \Gamma \left(z_3+z_4\right)\Gamma \left(z_1+z_2+z_3+z_4+1\right)\times  \non\\&
\Gamma \left(-\epsilon -z_1-z_2-\tfrac{3}{2}\right) \Gamma \left(-\epsilon +z_2-\tfrac{1}{2}\right) \Gamma \left(-\epsilon -z_1-z_3-\tfrac{3}{2}\right) \Gamma \left(-\epsilon +z_3-\tfrac{1}{2}\right) \times \non\\ &
\frac{\Gamma \left(\epsilon +z_1-z_4+\frac{5}{2}\right) \Gamma \left(\epsilon +z_4+\frac{3}{2}\right)\,\,x^{-z_1}}
{\Gamma \left(z_2+z_4+1\right) \Gamma \left(z_3+z_4+1\right) \Gamma \left(-2 \epsilon +z_1-z_4\right)}
\end{align}
The complex contour of integration in this four--fold Mellin--Barnes integral takes all variables in a straight line from $-i\infty$ to $i\infty$ with possible indentations, if necessary in order to leave the whole series of poles coming from Gamma functions of the form $\Gamma(...+z)$ to the left of the contour and the whole series of poles coming from Gamma functions of the form $\Gamma(...-z)$ to the right. Note that the contour of integration is well defined thanks to the presence of the $\e$ regulator inside the Gamma functions since it separates left and right poles which would otherwise collide.

Since there is a $\Gamma(-1\!-2\epsilon)$ in the denominator of (\ref{MBRep}), the expression could seem of order $\e$ at first sight; this is a common feature in the Mellin--Barnes representation of Feynman integrals. Actually, due the phenomenon of colliding poles in the $\e\rightarrow 0$ limit, the integral itself has poles in $\e$ producing a non vanishing result. The way of proceeding is to analyze case by case where this situation is produced and deforming the contour by picking up residues so as to avoid potentially colliding poles in the $\e\rightarrow 0$ expansion. Thanks to this method, the Mellin--Barnes integral becomes effectively at most two--fold, since we discard $\mathcal{O}(\e)$ terms. This procedure is explained in full detail in the literature (see for instance \cite{Smirnov}) and has been successfully applied to many four dimensional examples \cite{Smirnov:1999gc}. In this case we employed the package {\tt MB.m} \cite{MB}, to automatically perform the $\e$ expansion and get a list of Mellin--Barnes integrals whose integrands are independent of the regularization parameter. From this analysis, one can infer the functional form of the result up to order $\e^0$
\begin{align}\label{eq:functional}
\mathcal{I}_{4s}^{(2)} = -\frac{\pi\, e^{2\g_E\e}}{(-s)^{4+2\e}}&\left[  \frac{1}{\e^2}\left(a_1\, x^2+a_2\, x\right)+\frac{1}{\e}\left(
\left(a_3\, x^2+a_4\, x\right)\log(x)+a_5\, x^2+a_6\, x\right) + \right.\nonumber\\
 & + \left(a_7\, x^2+a_8\, x\right)\log(x)+a_9\, x^2+a_{10}\, x\bigg] + {\cal O}(\e)
\end{align}
where again $x=s/t$ is the ratio between the Mandelstam kinematic variables and the $\p$ factor has been collected for future convenience. It is straightforward to derive this functional dependence on $x$, since due to the procedure of picking up poles in $\e$ described above, there is no explicit dependence on $x$ in the Mellin--Barnes integrals, which therefore only contribute to give numerical coefficients, but not functions of the kinematic variables. The nontrivial logarithmic dependence of the result is indeed produced just by the $\e$ expansion.

After the expansion is performed, we obtain a total of $108$ one-fold and two-fold integrals (a list may be found in the attached {\tt Mathematica} notebook) needed to get the values of the coefficients ${a_i}$.
This computation is a little bit cumbersome, due to the large amount of integrals, but in every case, after appropriate manipulations, they can all be reduced to a form where some corollary of the Barnes' first and second lemmas may be applied. Almost all of the $108$ integrals could be solved with the extensive list of Barnes' lemmas corollaries detailed in \cite{Smirnov}. Those which were not present in that list, are explained in Appendix \ref{app:MB}.

Following this procedure we analytically determined the value of the coefficients to be
\begin{gather}
a_1= 6 \quad,\quad a_2 = 4 \quad \,, \quad a_3 = 6 \quad,\quad a_4 = 4,  \non\\
a_5= 20-12\gamma_E+12\log 2 \quad \!\!, \quad a_6=27-8\gamma_E+8\log 2, \non\\
a_7= 28 - 12\gamma_E \quad,\quad a_8= 28 - 8\gamma_E , \non\\
a_9 = -28-40 \gamma_E +12 \gamma_E ^2-8 \pi ^2-24 \log^2 2+80 \log 2 - 24 \gamma_E  \log 2,
\non\\
a_{10} = 46-54 \gamma_E +8 \gamma_E ^2-\frac{16 \pi ^2}{3}-16 \log^2 2+52 \log 2 - 16 \gamma_E  \log 2
\end{gather}
Plugging these into (\ref{eq:functional}) one gets the three dimensional scalar doublebox (in the $s$--channel) up to ${\cal O}(\e)$.

In order to compute the whole four--point amplitude, we use the four dimensional result \cite{Bern:1997nh} and plug the three dimensional integrals we compute
\begin{equation}
{\cal M}_4^{(2)} = \frac14\, st \left( s\, \mathcal{I}_{4s}^{(2)} +  t\, \mathcal{I}_{4t}^{(2)} \right)
\end{equation}
where the latter integral can be obtained from the former by replacing $s \leftrightarrow t$.
After some algebra and suitably redefining the mass scale $\mu$ as
\begin{equation}\label{eq:scheme}
\mu^2 = \mu'^2 \exp \left(\frac{47}{20} + \log 2 \right)
\end{equation}
we obtain the form of the two--loop $3d$ ${\cal N}=8$ four--point amplitude, which constitutes the main result of the paper
\begin{equation}\label{eq:result}
\boxed{
\begin{array}{ll}
& \\[-2mm]
~ {\cal M}_4^{(2)} = 5\p\, \displaystyle \frac{{\bf s}+{\bf t}}{{\bf st}}\, & \left[-\displaystyle  \frac{\left(\displaystyle -{\bf s}/\mu'\right)^{-2\e}}{(2\,\e)^2} - \frac{\left(\displaystyle -{\bf t}/\mu'\right)^{-2\e}}{(2\,\e)^2}
+ \frac12\, \log ^2\left(\displaystyle \frac{\bf s}{\bf t}\right)
+ 4\, \zeta_2 + 3 \log^2 2 \right. ~ \\[7mm]
& \left. ~~ + \displaystyle \frac15\left(\log 2 - \tfrac74\right)\, \displaystyle \frac{{\bf s}-{\bf t}}{{\bf s}+{\bf t}}\, \log \left(\displaystyle \frac{\bf s}{\bf t}\right)
- \frac{19}{10} \log 2 + \displaystyle \left(\frac{43}{20}\right)^2
\right]\\[-4mm]
 & 
\end{array}
}
\end{equation}

\section{Properties of the result.}\label{sec:properties}

We comment some relevant properties of the result.
First we note that since ${\cal M}$ is the coefficient in the perturbative series of powers in the coupling constant $g^2_{YM}$, which has mass dimension 1, the two--loop ratio is dimensionful as well.
The overall scale $\left( \tfrac{1}{s} + \tfrac{1}{t} \right)$ presents single poles in the collinear regimes $s \rightarrow 0$ and $t \rightarrow 0$, in contrast with ABJM \cite{CH,BLMPS}.

Second we verify that infrared divergences appear at two loops. These are captured by the $1/\e^2$ poles, whereas the subleading pole may be non--trivially reabsorbed by the scheme change (\ref{eq:scheme}). As expected from the planar limit, these divergences come from an $s$ channel and a $t$ channel contributions. After reabsorbing the $1/\e$ pole the finite piece of the amplitude becomes independent of the $\mu$ regularization scale.

\paragraph{Comparison to Wilson loops.}
As the theory obviously lacks both ordinary and dual conformal invariance, one should not expect a WL/amplitude duality to work.
To ascertain this, we compute a four cusp light--like Wilson loop at one loop. The necessary tools for this are given in Appendix \ref{app:WL}. In order to get a real result we restrict to a light--like polygonal contour with space--like non--vanishing invariants (the diagonals), and set $s\equiv x_{13}^2 < 0$ and $t\equiv x_{24}^2 < 0$. We also remark that at one loop no UV divergences arise, so that the calculation may be safely performed in $d=3$ dimensions.
Finally the expectation value of this Wilson loop reads
\begin{equation}
\langle W_4 \rangle = 1 - \frac{g^2_{YM}C_F}{\p}\, \frac{\sqrt{s\, t}}{\sqrt{-s-t}}\, {\rm ArcTanh} \left( \frac{\sqrt{s\, (s+t)} + \sqrt{t\, (s+t)}}{\sqrt{s\, t} - s - t} \right) + {\cal O}(g_{YM}^4) 
\end{equation}
where $C_F = T^a T^a$ is the quadratic Casimir of $U(N)$, which is $N/2$ in the fundamental representation.
Comparing it with the amplitude result, which vanishes at this order, we verify that the duality already fails in the simplest case.

Not even the UV divergent piece of the light--like Wilson loop, which appears first at two loops, resembles the IR divergences of the amplitude.
We checked this by analyzing planar corrections to a Wilson light--like line cusp at second order in perturbation theory, from which the divergences of the polygonal Wilson loop arise.
The computation parallels the four dimensional one and we won't go through it, but just state the result. The relevant tools for it to be performed may be found in Appendix \ref{app:WL}.
We verified that by virtue of the milder UV behavior of the gluon $x$--space propagator (\ref{eq:prop}) in three dimensions, where YM theory is superrenormalizable, the rainbow diagram\footnote{The one obtained by expanding the Wilson loop exponential to fourth order and Wick contracting two pairs of gluons lying on opposite edges of the cusp in a planar way.} as well as those involving the cubic gluon interaction are finite. The only source of divergence comes from the gluon self energy insertion.
The one--loop corrected propagator (\ref{eq:prop1}) is indeed finite in momentum space.
However, on dimensional grounds, its dependence on $p^2$ is through $(-p^2)^{-3/2-\e}$. In dimensional regularization, its Fourier transform develops a single pole in $\e$, which accounts for the UV divergence.
In fact, once the corrected propagator is inserted in the Wilson loop, no new singularities appear and the divergent piece of the Wilson loop reads
\begin{equation}
\langle W_4 \rangle ^{(2)} \big|_{UV} = - \frac{g^4_{YM}C_F N}{2^5\pi^2}\, \frac{1}{\e}\, \left(-s - t \right)
\end{equation}
where $s$ and $t$ stand again for the diagonals of the WL contour.

\paragraph{Transcendentality.}
We observe that the result does not respect maximal transcendentality, in contrast with ABJM.
Once the integral is normalized as in (\ref{eq:intnorma}), all $\g_E$ factors disappear, although leaving coefficients with mixed transcendentality, such as the one multiplying the $\log s/t$ piece, or the constant. Still, we point out that the maximal transcendentality piece of the constant coincides with that of ABJM.

\section{Comparison to ABJM: the Regge limit.}

We want to make a comparison of the ${\cal N}=8$ SYM two--loop four--point ratio (\ref{eq:result}) with the same result for ABJM, which we recall
\begin{equation}\label{eq:abjm}
{\cal M}_4^{(2)\,ABJM} =  -  \frac{\left(-s\right)^{-2\e}}{(2\,\e)^2} - \frac{\left(-t\right)^{-2\e}}{(2\,\e)^2} + \frac12\, \log ^2\left(\frac{t}{s}\right)
+ 4\, \zeta_2 + 3 \log^2 2
\end{equation}
The form of this amplitude coincide, up to rescalings and up to the constant, with the one--loop amplitude of ${\cal N}=4$ SYM in four dimensions
\begin{equation}
{\cal M}_4^{(1)\,{\cal N}=4} =  - \frac{\left(-s\right)^{-\e}}{\e^2} - \frac{\left(-t\right)^{-\e}}{\e^2} + \frac12\, \log ^2\left(\frac{t}{s}\right)
+ 4\, \zeta_2 
\end{equation}
Comparing (\ref{eq:result}) with (\ref{eq:abjm}), we find the following relation
\begin{equation}
\pi^{-1}\, \frac{st}{s+t}\, {\cal M}_4^{(2)\,{\cal N}=8} = 5\, \, {\cal M}_4^{(2)\,ABJM} + \left(\log 2 - \displaystyle \frac74\right)\, \frac{s-t}{s+t}\, \log \frac{s}{t} + C + {\cal O}(\e)
\end{equation}
Curiously enough, written in this form, the part of the constant of the ${\cal N}=8$ SYM result with maximal transcendentality coincides with the ABJM one and the difference are terms with lower transcendentality, namely
\begin{equation}
C = - \frac{19}{2} \log 2 + \frac{43^2}{80}
\end{equation}
The relevant part of the amplitude which differs from the ABJM result is the extra piece proportional to the logarithm of the ratio of the Mandelstam variables.
This suggests that we might compare the ${\cal N}=8$ result with that of ABJM in the Regge limit ${\bf t}/{\bf s}\ll 1$, where the leading logarithm approximation is pursued.
To make the comparison sensible, we need to deal with objects of the same dimension.
While the perturbative expansion of ABJM amplitudes is carried out in terms of the dimensionless 't Hooft parameter $\lambda_{CS}=N/k$, the Yang--Mills effective coupling constant in the planar limit is $\lambda_{YM}=g^2_{YM}N$, which in three dimensions is dimensionful, having mass dimension one.
While the Chern--Simons coupling may be tuned to be small adjusting the rank and the level in such a way that $N\ll k$, in SYM the coupling constant has to be compared to another energy scale.
In the four--point amplitude there are two characteristic energies of the process associated to $t$ and $s$ Mandelstam variables. Therefore perturbation theory is valid whenever $g^2_{YM} \ll \sqrt{s},\sqrt{t}$.
We may further fix $t$ to be the energy scale we are interested in, and study the small $t/s$ regime.
In such a limit we can rescale the Yang--Mills coupling $g^2_{YM}=g_{eff}^2\,\sqrt{t}$, defining a dimensionless effective coupling $g_{eff}^2=g_{YM}^2/\sqrt{t}$, which we take to be small ${g_{eff}}^2\ll 1$. In this way we establish the hierarchy of energy scales we are interested in, which is $g_{YM}^2\ll \sqrt{t}\ll \sqrt{s}$.
In addition, to make the comparison more direct, we can focus on the finite pieces of the SYM and ABJM amplitudes, rescaling the latter by the $5\pi$ factor, such that the normalization of the infrared divergences coincides between the two.
Having done this, and in terms of the dimensionless coupling, the $\mathcal{N}=8$ amplitude becomes
\begin{align}
{\cal M}_4^{(2)\,{\cal N}=8}\big|_{finite} = ~ & 
(1+{\bf y})\, \bigg[ \frac12\, \log ^2 {\bf y}
+ 4\, \zeta_2 + 3 \log^2 2 \nonumber\\
\label{YMRegge}
& ~~~~~~~~~~~ \left. + \frac15\left(\log 2 - \tfrac74\right)\, \frac{{\bf y}-{1}}{{\bf y}+{1}}\, \log {\bf y}
- \tfrac{19}{10} \log 2 + \left(\tfrac{43}{20}\right)^2
\right]
\end{align}
where ${\bf y}=1/{\bf x}={\bf t}/{\bf s}$ and $\lambda_{eff} = g_{eff}^2 N$.
\FIGURE{
  \centering
\includegraphics[width = .59 \textwidth]{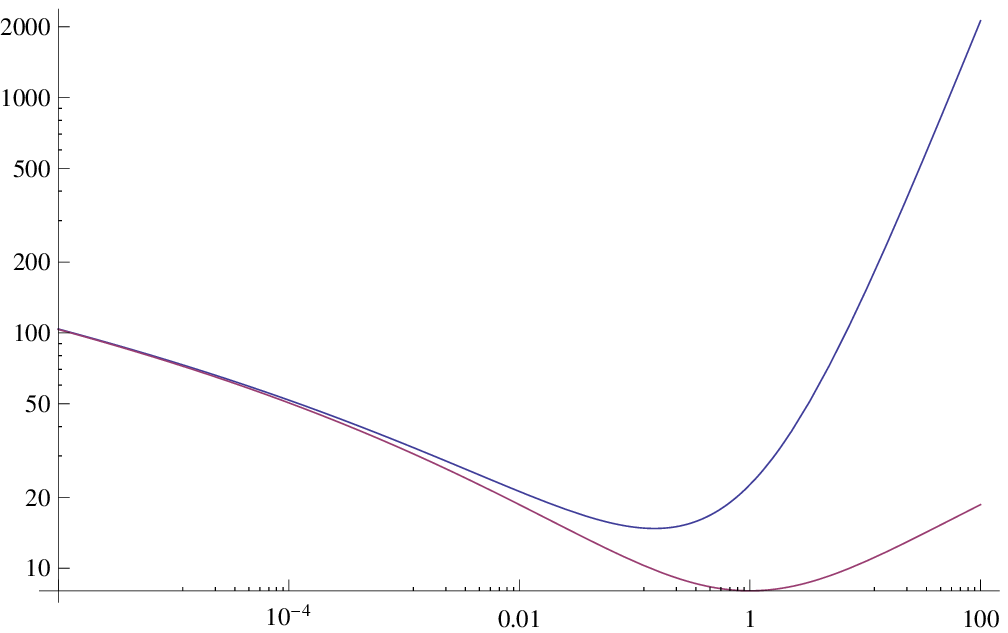}
\caption{Log-Log plot of $\mathcal{N}=8$ and ABJM amplitudes as functions of ${\bf y}={\bf t}/{\bf s}$.}
\label{fig:regge}
}

In the Regge limit ${\bf y}\ll 1$ this becomes
\begin{equation}
{\cal M}_4^{(2)\,{\cal N}=8}\big|_{finite} \xrightarrow{{\bf y}\ll 1} 
\frac12\, \log ^2 {\bf y}+\mathcal{O}(\log{\bf y})
\end{equation}
while ABJM, in the same limit becomes
\begin{equation}
{\cal M}_4^{(2)\,ABJM}\big|_{finite} \xrightarrow{{\bf y}\ll 1} 
\frac12\, \log ^2 {\bf y}+\mathcal{O}(1)
\end{equation}
Ergo, the finite piece of both amplitudes coincide in the Regge limit if we identify their dimensionless 't Hooft couplings $\lambda^2_{CS}\leftrightarrow \lambda^2_{eff}$. In Fig. \ref{fig:regge} we plot equation (\ref{YMRegge}) and the finite piece of (\ref{eq:abjm}) as functions of ${\bf y}={\bf t}/{\bf s}$ in logarithmic scale to illustrate the coincident Regge limit for small ${\bf t}/{\bf s}$.

Amplitudes in ${\cal N}=6$ CSM theory exhibit dual conformal invariance and, in the four point case, duality to the light--like four--polygonal Wilson loop. This has dramatic consequences on the dependence of the amplitude on the kinematic invariants, which has to satisfy an anomalous conformal Ward identity.
In the four--point case, at two loops, this fixes the finite piece of the amplitude to have a $\log^2 s/t$ form and its coefficient to be half that governing the IR singularities.
It is thus very pleasing that in Regge limit the leading term of the finite piece of the MSYM amplitude is exactly $\log^2 s/t$, and its coefficient of the form dictated by the anomalous conformal Ward identity.

\vskip 25pt
\section*{Acknowledgements}
\noindent

We would like to thank Arthur Lipstein for interesting comments on this work.

We are also grateful to Silvia Penati and Eoin O'Colgain for helpful suggestions on the draft of this paper.

Part of this work was performed at the Corfu Summer Institute 2012. We want to thank the organizers for providing the stimulating environment.
ML would like to thank the Department of Physics at the University of Milano--Bicocca for its kind hospitality during the completion of this work.

We also thank Federico Bonetti, Luca Giacone, Alexander Haupt, Claudius Dieter Reinhard Klare and Marta Leoni for many inspiring comments during our stay in Corfu.

The work of MSB is supported by the research grants MICINN-09-FPA2009- 07122 and MEC-DGI-CSD2007-00042.
The work of ML is supported by the research project CONICET PIP0396.

\vfill
\newpage
\appendix

\section{Tools for the WL computation.}\label{app:WL}

We work in three dimensional Minkowski spacetime with metric $g_{\mu\nu}=diag\left( 1, -1, -1 \right)$.
We use dimensional regularization of momentum integrals $d=3-2\e$, and introduce a mass scale $\mu$ to keep the action dimensionless.
The Wilson line operator is defined as
\begin{equation}
W[{\cal C}] = \frac{1}{N}\, {\rm Tr}\, {\cal P}\, e^{i\, g\, \int_{\cal C} dx^{\mu}\, A_{\mu}(x)}
\end{equation}
The $x$--space propagator for the gluon in $3d$ SYM, in dimensional regularization and in the Feynman gauge, reads
\begin{equation}\label{eq:prop}
\langle A_{\mu}(x)\, A_{\nu}(0) \rangle \equiv G_{\mu\nu}^{(0)}(x) = - \frac{\Gamma(\tfrac12-\e)\, \mu^{2\e}}{4\pi^{\frac32 - \e}}\, \frac{g_{\mu\nu}}{\left(-x^2\right)^{1/2-\e}}
\end{equation}
The momentum space one--loop corrected gauge propagator reads
\begin{equation}
\langle A_{\mu}(p)\, A_{\nu}(-p) \rangle ^{(1)} \equiv G_{\mu\nu}^{(1)}(p) =
\left( \frac{-i}{p^2} \right)\, \P_{\mu\nu}^{(1)}(p)\, \left( \frac{-i}{p^2} \right)
\end{equation}
where the gluon polarization operator evaluates, in dimensional regularization,
\begin{equation}
\P_{\mu\nu}^{(1)}(p) = i g^2 \frac{\Gamma\left( \tfrac32-\e \right)\Gamma\left( \tfrac12-\e \right)\Gamma\left( \tfrac12+\e \right)}{(4\pi)^{3/2-\e}\, \mu^{-2\e}\, \Gamma\left( 3-2\e \right)}
\left( (7-6\e) N - (2-4\e) n_f - n_s  \right) \frac{ p^2 g_{\mu\nu} - p_{\mu}p_{\nu}}{(-p^2)^{1/2+\e}}
\end{equation}
For ${\cal N}=8$ SYM, where $n_f = 4 N$ and $n_s = 7 N$,
\begin{equation}
G_{\mu\nu}^{(1)}(p) =
i g^2\, N\, \mu^{2\e}\, \frac{\Gamma\left( \tfrac32-\e \right)\Gamma\left( \tfrac12-\e \right)\Gamma\left( \tfrac12+\e \right)}{(4\pi)^{3/2-\e}\, \Gamma\left( 3-2\e \right)}\,
\left( -8+10\e  \right)\, \frac{1}{(-p^2)^{3/2+\e}}\, \left[g_{\mu\nu} - \frac{p_{\mu}p_{\nu}}{p^2}\right]
\end{equation}
and Fourier transforming to $x$--space with
\begin{equation}
\int \frac{d^d p}{(2\pi)^d}\, \frac{e^{-ipx}}{(-p^2)^a} = \frac{i}{4^a\, \pi^{d/2}}\, \frac{\Gamma\left(\tfrac{d}{2}-a\right)}{\Gamma\left(a\right)}\, (-x^2)^{a-d/2}
\end{equation}
this becomes
\begin{align}\label{eq:prop1}
D_{\mu\nu}^{(1)} &= \mu^{2\e}\, \int \frac{d^{3-2\e} p}{(2\pi)^{3-2\e}}\, e^{-ipx}\, G_{\mu\nu}^{(1)}(p) = \non\\
& = - g^2\, N\, \left(\mu^{2\e}\right)^2\,
\frac{\Gamma\left( \tfrac32-\e \right)\Gamma\left( \tfrac12-\e \right)\Gamma\left( \tfrac12+\e \right)}{2^6\, \pi^{3-2\e}\, \Gamma\left( 3-2\e \right)}\,
\left( -8+10\e  \right)\, \times \non\\&
\left[
\frac{\Gamma\left( -2\e \right)}{\Gamma\left( \tfrac32+\e \right)}
(-x^2)^{2\e}\, g_{\mu\nu}
- \frac14\, \frac{\Gamma\left( -1-2\e \right)}{\Gamma\left( \tfrac52+\e \right)}\,
\partial_{\mu} \partial_{\nu}\, (-x^2)^{1+2\e} \right]
\end{align}

\section{New corollaries.}\label{app:MB}

To complete the computation we also had to derive two corollaries of Barnes' first lemma, to solve the integrals (in the first $\l_1+\l_3\neq 0$ is assumed)
\begin{align}
&\frac{1}{2\p i}\, \int_{-i\infty}^{i\infty}\, dz\, \Gamma \left(z+\lambda _1\right) \Gamma \left(z+\lambda _2\right) \Gamma^{**} \left(-z-\lambda _2-1\right) \Gamma \left(\lambda _3-z\right) \psi ^{(0)}\left(-z-\lambda _2-1\right)\, , \non\\&
\frac{1}{2\p i}\, \int_{-i\infty}^{i\infty}\, dz\, \Gamma \left(z-\lambda _1+1\right) \Gamma^* \left(z-\lambda _2-1\right) \Gamma^* \left(\lambda _2-z\right) \Gamma \left(\lambda _1-z\right) \psi ^{(0)}\left(\lambda _2-z\right) \non
\end{align}
These may be found from lemmas (D.12) and (D.14) of \cite{Smirnov}. To do this one may first insert a fictitious regulator to avoid coincident left and right poles. Having done this one takes a residue to shift the contour in such a way to remove one asterisk from the proper Gamma function and bring the integral to a form where the aforementioned corollaries of first Barnes' lemma may be applied. Since the original integral was well--defined, the poles in the regulator disappear in the sum of the two pieces and the limit where it approaches zero can be safely taken, to get the final result.
We thus obtain
\begin{align}
& \frac{1}{2\p i}\, \int_{-i\infty}^{i\infty}\, dz\, \Gamma \left(z+\lambda _1\right) \Gamma \left(z+\lambda _2\right) \Gamma^{**} \left(-z-\lambda _2-1\right) \Gamma \left(\lambda _3-z\right) \psi ^{(0)}\left(-z-\lambda _2-1\right)
= \non\\ &
\frac{1}{12 \Gamma \left(\lambda _1+\lambda _3-1\right)}
\big[\Gamma \left(\lambda _1-\lambda _2\right) \Gamma \left(\lambda _1+\lambda _3-1\right) \Gamma \left(\lambda _2+\lambda _3\right) \times \non\\&
\left( -12 \gamma  \psi ^{(0)}\left(\lambda _1-\lambda _2\right)+12 \gamma  \psi ^{(0)}\left(\lambda _2+\lambda _3\right)+\pi ^2+6 \gamma ^2\right) +
\non\\&
+\Gamma \left(\lambda _1-\lambda _2-1\right) \left(\Gamma \left(\lambda _1+\lambda _3-1\right) \Gamma \left(\lambda _2+\lambda _3+1\right) \left(-12 (\gamma -1) \psi ^{(0)}\left(\lambda _1-\lambda _2-1\right) +
\right.\right. \non\\& \left.
+12 (\gamma -1) \psi ^{(0)}\left(\lambda _2+\lambda _3+1\right)+\pi ^2+6 \gamma ^2-12 \gamma +12\right)-\Gamma \left(\lambda _1+\lambda _3\right) \Gamma \left(\lambda _2+\lambda _3\right) \times
\non\\&
 \left(6 \psi ^{(0)}\left(\lambda _1+\lambda _3-1\right){}^2+12 (\gamma -1) \psi ^{(0)}\left(\lambda _1+\lambda _3-1\right)-6 \psi ^{(0)}\left(\lambda _2+\lambda _3\right){}^2 + \right.
\non\\&
 -12 \psi ^{(0)}\left(\lambda _1-\lambda _2-1\right) \left(\psi ^{(0)}\left(\lambda _1+\lambda _3-1\right)-\psi ^{(0)}\left(\lambda _2+\lambda _3\right)+\gamma -1\right) +
 \non\\& \left. \left.\left.
 -6 \psi ^{(1)}\left(\lambda _1+\lambda _3-1\right)-6 \psi ^{(1)}\left(\lambda _2+\lambda _3\right)+\pi ^2+6 \gamma ^2-12 \gamma +12\right)\right)\right]
\end{align}
\begin{align}
& \frac{1}{2\p i}\, \int_{-i\infty}^{i\infty}\, dz\, \Gamma \left(z-\lambda _1+1\right) \Gamma^* \left(z-\lambda _2-1\right) \Gamma^* \left(\lambda _2-z\right) \Gamma \left(\lambda _1-z\right) \psi ^{(0)}\left(\lambda _2-z\right)
 =
\non\\&
\frac{1}{12} \big[\Gamma \left(\lambda _1-\lambda _2\right) \Gamma \left(-\lambda _1+\lambda _2+1\right) \times
\non\\ &
\left(12 (\gamma -1) \psi ^{(0)}\left(\lambda _1-\lambda _2\right)-12 (\gamma -1) \psi ^{(0)}\left(-\lambda _1+\lambda _2+1\right)+\pi ^2+6 \gamma ^2-12 \gamma +12\right) +
\non\\&
-\Gamma \left(\lambda _1-\lambda _2-1\right) \left(12 \Gamma \left(-\lambda _1+\lambda _2+1\right) \left(\psi ^{(0)}\left(-\lambda _1+\lambda _2+1\right)+1\right)+
\right. \non\\&
+\Gamma \left(-\lambda _1+\lambda _2+2\right) \left(6 \psi ^{(0)}\left(\lambda _1-\lambda _2-1\right){}^2-12 \psi ^{(0)}\left(-\lambda _1+\lambda _2+2\right) \psi ^{(0)}\left(\lambda _1-\lambda _2-1\right)+
\right.
\non\\&
+6 \psi ^{(0)}\left(-\lambda _1+\lambda _2+2\right){}^2+6 \psi ^{(1)}\left(\lambda _1-\lambda _2-1\right)+6 \psi ^{(1)}\left(-\lambda _1+\lambda _2+2\right) +
\non\\& \left.\left.\left.
-\pi ^2-6 \gamma ^2+12 \gamma -12\right)\right)\right]
\end{align}

\end{document}